\documentclass[twocolumn,showpacs,preprintnumbers,amsmath,amssymb]{revtex4}
\usepackage{graphicx}
\usepackage{bm}
\usepackage{color}

\newcommand{ \be }{\begin{equation}}
\newcommand{ \ee }{\end{equation}}
\newcommand{ \bea }{\begin{eqnarray}}
\newcommand{ \eea }{\end{eqnarray}}

\newcommand{ \ds }{\displaystyle}
\renewcommand{\ol}[1]{\overline{#1}}
\newcommand{ \mean }[1]{\left< #1 \right>}

\newcommand {\psirp} {\Psi_{\rm RP}}
\newcommand {\psinep} {\Psi_{\rm EP}^n}

\newcommand {\mxnrp} {\mean{x_n}_{\psirp}}
\newcommand {\mynrp} {\mean{y_n}_{\psirp}}

\newcommand {\cn}[1]{\overline{c}_{#1}}
\newcommand {\sn}[1]{\overline{s}_{#1}}
\newcommand {\an}[2]{1 #1 \overline{c}_{2 #2}}

\newcommand {\lc}[3]{  \frac{v_{#1}}{v_{#2}}\frac{\cn{#1\mp#2}}{a^{#3}_{2#2}} }
\newcommand {\ls}[3]{  \frac{v_{#1}}{v_{#2}} \frac{\sn{#1\mp#2}}{a^{#3}_{2#2}} }

\begin{document}

\title{Effects of non-uniform acceptance in anisotropic flow measurement}

\author{Ilya Selyuzhenkov and Sergei Voloshin}
\affiliation{Wayne State University, Detroit, Michigan, 48201}
\date{\today}

\begin{abstract}
Applicability of anisotropic flow measurement techniques and their 
extension for detectors with non-uniform azimuthal acceptance are discussed.
Considering anisotropic flow measurement with
two and three (mixed harmonic) azimuthal correlations
we introduce a set of observables based on the $x$ and $y$ components of 
the event flow vector.
These observables provide independent measures of anisotropic flow, and
can be used to test self-consistency of the analysis.
Based on  these observables we propose a technique that explicitly 
takes into account the effects of non-uniform detector acceptance.
Within this approach the acceptance corrections, as well as
parameters which define the method applicability, can be determined directly 
from experimental data.
For practical purposes a brief summary of the method is provided at the end.
\end{abstract}



\maketitle


\section{\label{Introduction} Introduction}

One of the most important observable in the study of 
ultra-relativistic nucleus-nucleus collisions is
the anisotropic transverse
flow~\cite{Back:2004je, Adams:2005dq, Adcox:2004mh, Voloshin:2002wa}.
It is quantified by coefficients in Fourier decomposition
of particle azimuthal distributions
relative to the collision reaction
plane~\cite{Voloshin:1994mz, Poskanzer:1998yz},
the latter is defined by the beam direction and the impact parameter.
The second harmonic in such a decomposition is called the elliptic flow.
Observation of strong 
in-plane elliptic flow increasing with collision energy  
from top AGS (Alternating Gradient Synchrotron)
energies~\cite{Barrette:1994xr, Barrette:1996rs}, 
then at CERN SPS 
(Super Proton Synchrotron)~\cite{Appelshauser:1997dg, Alt:2003ab}, and 
recently at RHIC (Relativistic Heavy Ion Collider)~%
\cite{Ackermann:2000tr, Adler:2001nb, Adler:2003kt, 
Adcox:2002ms, Back:2002gz, Back:2004mh, Back:2004zg}
shows that the bulk matter created in the high energy heavy 
ion collision strongly
interacts and behaves collectively.
Taken together with a key feature of anisotropic flow to be sensitive
to the early stage of the collision, this indicates rapid
attainment of local thermal equilibrium in the created system.
For central collisions, RHIC results 
are in agreement with ideal (zero viscosity) hydrodynamic 
predictions~\cite{Ollitrault:1992bk, Teaney:2001gc, Teaney:2001av}.
These observations have played an important role in
the discovery of the
strongly interacting Quark Gluon Plasma (sQGP),
the new form of matter formed in heavy ion collisions.

There exist a set of different techniques for anisotropic 
transverse flow measurement,
which have been successfully applied at a variety of
experimental setups worldwide~%
\cite{Barrette:1994xr, Barrette:1996rs, Appelshauser:1997dg, 
Alt:2003ab, Ackermann:2000tr, Adler:2001nb, Adler:2003kt,
Adcox:2002ms, Back:2002gz, Back:2004mh, Back:2004zg}.
As the reaction plane is not known experimentally, various methods 
exploit azimuthal correlations between particles as all of
the particles are correlated 
to the same reaction plane.
The most widely used are
the event plane~\cite{ Barrette:1996rs, Poskanzer:1998yz},
the scalar product~\cite{Poskanzer:1998lbnl, Adler:2002pu},
and mixed harmonic~\cite{Poskanzer:1998yz, Borghini:2002vp,
Adams:2003zg, Adams:2004bi}
methods.
Analysis of anisotropic flow with azimuthal correlations
requires to examine contribution from effects
not related to orientation of the reaction plane,
such as resonance decays, jets,
as well as effects of conservation laws (charge, momentum, etc).
Most of these so called {\it non-flow} correlations are 
due to few particle interactions,
and their relative contribution scales
inversely proportional to particle multiplicity in the event.
For that reason in multi-particle correlations these effects are suppressed
compared to collective effects such as anisotropic flow.
The complete consideration of methods to study non-flow effects
is beyond the scope of this paper
(for more discussions see, for example,~\cite{Adler:2002pu} 
and references therein).
In this paper we assume that the azimuthal distributions
of particles produced in the collision depends only
on the orientation relative to the reaction plane.

High statistics experimental data collected in recent years at
RHIC allow to perform very precise measurements~%
\cite{Barrette:1994xr, Barrette:1996rs, Appelshauser:1997dg,
Alt:2003ab, Ackermann:2000tr, Adler:2001nb, Adler:2003kt,
Adcox:2002ms, Back:2002gz, Back:2004mh, Back:2004zg}.
With the new data,
the systematic uncertainty in the analysis becomes dominant compared
to the statistical errors.
It is vital to carefully investigate the systematic errors, in
particular due to non-perfect azimuthal acceptance,
as well as to review the applicability of different methods in this respect.
In addition, the availability of different experimental
setups with complicated azimuthal acceptance,
such as central arms of PHENIX (A Physics Experiment at RHIC) 
detector~\cite{Aizawa:2003zq, Adcox:2003zp},
NA49 TPC (Time Projection Chamber)~\cite{Afanasev:1999iu},
and PHENIX/STAR (Solenoidal Tracker At RHIC) ZDC SMD
(Zero Degree Calorimeter Shower Maximum Detector)~\cite{ZDCSMDproposal},
requires further development of new and/or generalization of known
methods for use with detectors with significant acceptance non-uniformity.
Such generalization would allow to enrich already available data 
with anisotropic flow measurement results
from a wider range of experimental setups.
This further provides an opportunity for a comprehensive
comparison of available theoretical predictions against the experimental data.

In this paper we describe a procedure to broaden the 
applicability of known methods to measure anisotropic flow
for a range of detectors with non-uniform azimuthal acceptance.
Distinctive feature of the suggested approach is that
the acceptance corrections can be determined directly from experimental data.
This eliminates the need to perform time-consuming 
and model dependent
Monte-Carlo simulations of the detector acceptance and efficiency.
The main idea is demonstrated on an example of
two-particle correlations, but for completeness 
we also provide formulae for the case of three particle
correlations (mixed harmonic) technique, which, as discussed above
are less susceptible to non-flow effects.
We keep the discussion of more complicated three particle correlation case
in separate subsections, such that they can be easily skipped if the reader is
interested only in the main idea.

The paper is organized as follows.
In section \ref{Method} 
we review notations and definitions of
basic quantities used in anisotropic flow analysis.
We formulate them in a way that later
helps us to introduce new observables based on
different components of the event flow vector.
These observables provide independent measures of anisotropic flow,
and can be used to test self-consistency of the results.
In section \ref{acceptanceXYseparetely}  we discuss
the effects of non-uniform detector acceptance,
and describe the procedure of correcting the single particle
and event flow vector such that
observables derived for the perfect detector can be used.
We provide the smallness parameters
that can be used to quantify the range of applicability
of the proposed method.
These parameters can be estimated directly
from experimental data.
Finally, we summarize the method.

A similar problem of flow analysis with non-uniform acceptance detector
was discussed in \cite{Borghini:2001vi, Bhalerao:2003xf}
in the context of cumulant and Lee-Yang zeros analyses
based on the use of generating functions.
Though in some cases these techniques might yield to similar results,
our independent approach clarifies the nature
of the acceptance effects in flow studies,
and further provides the required tools for
the most often used analyses based on correlations with the event flow vector.


\section{Methods }
\label{Method}

\subsection{\label{Notations} 
Definition and notations}

Anisotropic transverse flow of particles produced in heavy ion collision
is quantified by coefficients in Fourier decomposition of
particle azimuthal distribution~\cite{Voloshin:1994mz, Poskanzer:1998yz}.
In this paper we use particle azimuthal spectra normalized to unity 
(particle production probability density):
\begin{eqnarray}
\label{FlowDefinition}
\rho(\phi-\psirp) 
= \frac{1}{2\pi}
  \left( 1+2\sum_{n=1}^{\infty}v_n \cos \left[ n \left(\phi - \Psi_{\rm
    RP}\right)\right] \right) .
\end{eqnarray}
Here $\phi$ is the particle azimuthal angle,
and $v_n$ is the $n$-th harmonic anisotropic flow.
The reaction plane angle $\psirp$ in equation~(\ref{FlowDefinition})
is an azimuthal angle of the impact parameter.
In general, coefficients $v_n$ are functions
of particle transverse momentum
$p_t$ and pseudorapidity $\eta$.
Such dependences are hereafter assumed implicitly and
not indicated in the notation for the sake of brevity and simplicity.

We introduce a unit vector $u_n$ ($n>0$), defined for each particle:
\begin{eqnarray}
\label{u_n}
u_n =  x_n + i y_n \equiv \cos n\phi + i \sin n\phi = \exp\{i n\phi\}.
\end{eqnarray}
In this equation $u_n$ is given
as a complex number with angle $n\phi$ and absolute value of unity.
Throughout this paper we use complex number representation,
but all equations can be re-written in terms of commonly used algebra of
2-dimensional vectors.
An estimate of the reaction plane orientation is usually
obtained with the so called $n$-harmonic
{\it event flow vector} $Q_n$, which 
is defined as a sum of $u_n$-vectors
over a specific subset "${\rm EP}$" of particles produced in the collision:
\begin{eqnarray}
\label{EPvector}
\nonumber Q_n &=& \sum_{\rm EP} u_n 
= \sum_{\rm EP}(\cos n \phi + i \sin n \phi)
\\ 
&\equiv& X_n + i Y_n = |Q_n|\exp\{i n\psinep\}.
\end{eqnarray}
Here $X_n$ and $Y_n$ are the event vector components,
and $\psinep$ is $n$-th harmonic {\it event plane angle}.
For each $n$ the event flow vector $Q_n$ (or $\psinep$)
provides an independent estimate of the reaction plane orientation.
Note that $u_n$ itself can serve as a $Q_n$-vector.
However this is not very practical, since the more particles used
to define the event plane
the closer its orientation will be to that of the reaction plane.

To derive the main formulae of proposed technique
we find it very useful to introduce,
though experimentally unobservable, $u_n$
averaged over events with fixed orientation of the reaction plane:
\begin{eqnarray}
\label{x_n_PSI}
\nonumber
\mean{u_n}_{\psirp} &=& \mxnrp + i\mynrp =
\int d\phi~u_n~ \rho (\phi-\psirp) 
\\
\nonumber
&
=& \int\frac{d\phi}{2\pi} ~u_n \left(1+2\sum_{m=1}^{\infty}
v_m\cos \left[ m \left(\phi - \psirp\right)\right]\right)
\\ 
&
=& v_n (\cos n \psirp + i \sin n \psirp).
\end{eqnarray}
In this section
we consider the case of a detector with perfect azimuthal acceptance.
This implies that an integration over azimuthal angle $\phi$ goes over
$2\pi$ without any weight.
We will relax this assumption in section \ref{acceptanceXYseparetely}
when considering detector acceptance effects.
Similarly:
\begin{eqnarray}
\label{Q_EP_vs_PSI}
\nonumber  \mean{Q_n}_{\psirp}
&=&\mean{X_n}_{\psirp} + i\mean{Y_n}_{\psirp}
\\
\nonumber 
&=& \mean{ M v_n}_{\rm EP} (\cos n \psirp 
+i  \sin n \psirp)
\\
&=&
 V_n (\cos n \psirp + i \sin n \psirp ).
\end{eqnarray}
Here $V_n \equiv  \mean{M v_n}_{\rm EP}$ is an average
$n$-harmonic anisotropic flow $v_n$ convoluted with multiplicity $M$
of particles from a subset ``EP'' used to calculate
the event flow vector $Q_n$.

From equation~(\ref{Q_EP_vs_PSI}) it follows
that $Q_n$ is defined as a vector in transverse plane,
which {\em on average} 
has an orientation of that of the reaction plane.
This feature can be used to define the event flow vector
with detectors without tracking that are only
sensitive to the shape of the particle distribution
in the transverse plane (for example, calorimeters).
The only, but very important, requirement to be fulfilled
is that the $Q_n$ components, $X_n$ and $Y_n$,
should be on average proportional to
$\cos n \psirp$ and $\sin n \psirp$, respectively.

\subsection{\label{MethodReview}
Anisotropic flow from different components}


\subsubsection{Two particle correlations}

Anisotropic flow via two particle correlations
can be obtained with the so-called {\it scalar product method}~%
\cite{Adler:2002pu}.
According to the scalar product technique,
one considers the average of the 
product of $u_n$ and $Q_n$ vectors over all events.
With the help of Eqs.~(\ref{x_n_PSI}) and (\ref{Q_EP_vs_PSI})
this average can be written as an average over
all events with fixed reaction plane with further average over all
reaction plane orientations:
\begin{eqnarray}
\label{meanDefin}
\mean{u_n Q_n^*} &=& \mean{ x_n X_n}  + \mean{y_n Y_n}
\\
&=& 
\nonumber
\ds \int\limits_{0}^{2\pi}\frac{d\psirp}{2\pi}
\mean{u_n}_{\psirp} 
\mean{Q_n^{*}}_{\psirp} = v_n V_n.
\end{eqnarray}
Here, angle brackets with subscripts $\psirp$,
$\mean{...}_{\psirp}$,
denote the average over events with fixed $\psirp$;
angle brackets without subscripts, $\mean{...}$, 
correspond to the average over entire ensemble of events with
all possible orientations of the reaction plane.

The left hand side of Eq.~(\ref{meanDefin})
can be measured from experimental data.
To obtain $v_n$
one needs to evaluate  $V_n$.
This can be done by using {\em random sub-events},
i.e. randomly assigning particles
used to construct the event flow vector
into two subsets $a$ and $b$~\cite{Poskanzer:1998yz}:
\begin{eqnarray}
\label{resolution}
\mean{ Q_n^a Q_n^{b*}} = \mean{X_n^a X_n^b}+  \mean{Y_n^a Y_n^b}= \frac{1}{4} V_n^2.
\end{eqnarray}
The factor of $1/4$ here takes into account
the multiplicity difference between the full event 
and that of sub-events $a$ and $b$.

From Eqs.~(\ref{meanDefin}) and (\ref{resolution}) we obtain:
\begin{eqnarray}
\label{v_n_sclalarProduct}
v_n = \frac{\mean{ u_n Q_n^*}}{2\sqrt{\mean{Q_n^a Q_n^{b*}}}}.
\end{eqnarray}
We further note that the two terms in
Eqs.~(\ref{meanDefin})
and (\ref{resolution}) 
are statistically independent, which allows to consider them separately:
\begin{eqnarray}
\label{ScalarObsXY}
\mean{x_n X_n}
 &=&  \mean{y_n Y_n} = \frac{1}{2} v_n V_n,
\\
\label{ScalarResXY}
\mean{X_n^a X_n^b} &=&  \mean{Y_n^a Y_n^b}= \frac{1}{8} V_n^2,
\end{eqnarray}
thus providing two independent measures of anisotropic flow:
\begin{eqnarray}
\label{XYseparetely}
v_n = \frac{\mean{x_n X_n}}{\sqrt{2 \mean{X_n^a X_n^b}}}
=\frac{\mean{y_n Y_n}}{ \sqrt{2\mean{Y_n^a Y_n^b}}}.
\end{eqnarray}
Independent observables (\ref{XYseparetely})
can be used to check the self-consistency of the results.

Note, that with normalization of the $Q_n$-vector to unity,
$Q_n \rightarrow Q_n/|Q_n|$,
the average $\mean{u_n Q_n^*}$ in (\ref{meanDefin}) reduces to
$\mean{\cos \left[ n (\phi -\psinep)\right]}$, and Eq. (\ref{v_n_sclalarProduct}) leads to
the main observable of the event plane method~\cite{Poskanzer:1998yz}:
\begin{eqnarray}
\label{standardMethod}
v_n = \frac{\mean{\cos\left[n(\phi-\Psi_{\rm EP}^{n;a,b})\right]}}
{\sqrt{\mean{\cos\left[ n(\Psi_{\rm EP}^{n;a} - \Psi_{\rm EP}^{n;b})\right]}}}.
\end{eqnarray}
Similarly,
the second equality in formula (\ref{XYseparetely}) gives an observable
used by the NA49 Collaboration \cite{Alt:2003ab}:
\begin{eqnarray}
\label{XYseparetelyNA49}
v_n = \sqrt{2}~\frac{\mean{\sin n\phi \cdot \sin n\Psi_{\rm EP}^{n; a,b}}}
{\sqrt{\mean{\sin n\Psi_{\rm EP}^{n;a} \cdot \sin n \Psi_{\rm EP}^{n;b}}}}.
\end{eqnarray}
Here and in Eq.~(\ref{standardMethod}) we use
the event plane angle defined for the subevents,
which resulted in an extra factor of two
compared to Eqs.~(\ref{v_n_sclalarProduct}, \ref{XYseparetely}).


\subsubsection{Three particle correlations}

In the case of three particle correlations one considers:
\begin{eqnarray}
\label{mixedHarmicCorrFunct}
&
\mean{ u^a_n u^b_n Q_{2n}^*} =
&
\nonumber
\\&
= \mean{x^a_n x^b_n X_{2n} - y^a_n y^b_n X_{2n} + x^a_n y^b_n Y_{2n} +y^a_n x^b_n Y_{2n}}
&\\
\nonumber &=
\int\limits_{0}^{2\pi}\frac{d\psirp}{2\pi}
\mean{u_n^a}_{\psirp} 
\mean{u_n^b}_{\psirp}
\mean{Q_{2n}^{*}}_{\psirp}
=v_n^2 V_{2n},
&
\end{eqnarray}
where:
\begin{eqnarray}
\label{resolutionMixed}
V_{2n}= \mean{M v_{2n}}_{\rm EP}
= 2\sqrt{\mean{Q_{2n}^a Q_{2n}^{b*}}} .
\end{eqnarray}
Then, the anisotropic flow $v_n$ is given by:
\begin{eqnarray}
\label{XYseparetelyMixed}
|v_n| = \sqrt{\frac{\mean{u_n^a u_n^b Q_{2n}}}
{2\sqrt{\mean{Q_{2n}^a Q_{2n}^{b*}}}}}.
\end{eqnarray}
If one normalizes $Q_n$-vector to unity, formula~(\ref{XYseparetelyMixed})
leads to an observable of the mixed harmonic method~\cite{Adams:2003zg, Adams:2004bi}:
\begin{eqnarray}
\label{mixedHarmonicPsi}
|v_n| = \sqrt{
\frac{\mean{\cos\left[n(\phi_a +\phi_b - 2\Psi_{\rm EP}^{2n;a,b})\right]}}
{\sqrt{\mean{\cos \left[2n(\Psi_{\rm EP}^{2n,a}-\Psi_{\rm EP}^{2n,b})\right]}}}}.
\end{eqnarray}
All terms in formula (\ref{mixedHarmicCorrFunct})
are statistically independent, which leads to a set of equalities:
\begin{eqnarray}
\label{mixedObsXY}
\nonumber & \mean{x_n^a x_n^b X_{2n}}
= - \mean{y_n^a y_n^b X_{2n}}
= \mean{x_n^a y_n^b Y_{2n}} = \mean{y_n^a x_n^b Y_{2n}}&
\\&= \frac{1}{4} v_n^2 V_{2n}.
\end{eqnarray}
Thus one obtains four independent
observables to measure anisotropic flow from three particle correlations:
\begin{eqnarray}
\label{MixedHarmonicXYseparetely}
\nonumber |v_n| & 
= & \sqrt{\sqrt{2}\frac{\mean{x_n^{a} x_n^{b} X_{2n}}}{\sqrt{\mean{X_{2n}^a X_{2n}^{b}}}}}
= \sqrt{- \sqrt{2}\frac{\mean{y_n^{a} y_n^{b} X_{2n}}}{\sqrt{\mean{X_{2n}^a X_{2n}^{b}}}}}
\\ &  = & \sqrt{\sqrt{2}\frac{\mean{x_n^{a} y_n^{b} Y_{2n}}}{\sqrt{\mean{Y_{2n}^a Y_{2n}^{b}}}}}
= \sqrt{\sqrt{2}\frac{\mean{y_n^{a} x_n^{b} Y_{2n}}}{\sqrt{\mean{Y_{2n}^a Y_{2n}^{b}}}}}.
\end{eqnarray}
As in the case of the two particle correlations,
each of the four terms in Equation~(\ref{MixedHarmonicXYseparetely})
provides an independent measure of anisotropic flow,
and can be used to check the self-consistency of the results.

Note, that one can construct three particle correlation function
from $u_n$ and $Q_m$ vector components other
than that defined by formula (\ref{mixedHarmicCorrFunct}).
Some examples are $\mean{u_n Q_n^a Q_{2n}^{b*} }$
or $\mean{Q_n^a Q_{n}^{b}u^*_{2n} }$.
Derivation of observables based on these correlators are similar, but
in this paper we {only consider} combination (\ref{mixedHarmicCorrFunct}),
which in the case of $n=1$ leads to the known observable for 
directed flow~\cite{Adams:2003zg, Adams:2004bi}.


\section{\label{acceptanceXYseparetely} Effects of non-uniform acceptance}
In order to generalize our consideration
for the case of imperfect acceptance
we introduce the acceptance function
$A(\phi)$ which we  normalize to unity
(similar to \cite{Borghini:2001vi, Bhalerao:2003xf}):
\begin{eqnarray}
\label{intA}
\int \frac{d\phi}{2\pi} A(\phi) = 1.
\end{eqnarray}
Then  the average of 
some function $f(\phi)$, which depends on particle azimuthal angle
$\phi$, at fixed reaction
plane orientation is given by the integral:
\begin{eqnarray}
\label{f_PSI_acc}
\mean{f}_{\psirp} 
&=& \int d\phi A(\phi) f(\phi) \rho (\phi-\psirp)
\\
\nonumber
&=& \overline{f} + 
2\sum_{m=1}^{\infty}v_m \left[\overline{f c}_m\cos m \psirp 
+ \overline{f s}_m\sin m \psirp\right].
\end{eqnarray}
Here for brevity we introduce notation $c_m=\cos m\phi$ and
$s_m=\sin m\phi$, and
denote by $\overline{f}$, the average over the detector acceptance:
\begin{eqnarray}
\label{averageDefinition}
\overline{f} = \int\frac{d\phi}{2\pi}~A(\phi) f(\phi).
\end{eqnarray}
One might note that $\cn{n}$ and $\sn{n}$ represent $n$-th
harmonic coefficients in the Fourier expansion of the acceptance function $A(\phi)$.
An important observation is that the acceptance average of $f$,
$\overline{f}$,
coincides with the event average, $\mean{f}$:
\begin{eqnarray}
\label{averagePhi}
\mean{f} &=& \frac{\int d\Psi_{RP}d\phi A(\phi) f(\phi) \rho(\phi-\psirp)}
{\int d\Psi_{RP}d\phi A(\phi) \rho(\phi-\psirp)} = \overline{f}.
\end{eqnarray}
We assume here  that the distribution of the reaction plane angle, $\psirp$,
is uniform within a given centrality event sample.
Experimentally, this can be achieved by using
for the collision centrality determination
the independent detector with a good azimuthal coverage.
Consequently, all acceptance average quantities
can be extracted directly from the experimental data by the
corresponding average over all particles in the event sample.

Formula~(\ref{f_PSI_acc})
allows to re-write the expressions for
$\mean{x_n}_{\psirp}$ and $\mean{y_n}_{\psirp}$
taking  into account the effects of the non-uniform detector acceptance.
For clarity of comparison with Eq.~(\ref{x_n_PSI}) we separate the term with $m=n$,
which is the only non-vanishing term in case of perfect acceptance:
\begin{eqnarray}
\label{x_n_PSI_acc_final}
\nonumber
\mean{x_n}_{\psirp} &=&
\overline{c}_n 
+ v_n a^{+}_{2n}
\left\{\cos n \psirp + \lambda^{s+}_{2n}\sin n \psirp \right.
\\\nonumber 
&&+ \sum_{m\ne n}^{\infty}
\left(
[\lambda^{c+}_{n-m} + \lambda^{c+}_{n+m}] \cos m \psirp
\right.
\\&&
\phantom{\sum_{m\ne n}^{\infty}}
+\left.\left.
[\lambda^{s+}_{n+m} - \lambda^{s+}_{n-m}] \sin m \psirp
\right) \right\},
\\
\label{y_n_PSI_acc_final}
\nonumber
\mean{y_n}_{\psirp} &=& \overline{s}_n
+ v_n a^{-}_{2n}
\left\{\sin n \psirp + \lambda^{s-}_{2n}\cos n \psirp \right.
\\
\nonumber
&& 
 + \sum_{m\ne n}^{\infty}
\left(
[\lambda^{c-}_{n-m}
- \lambda^{c-}_{n+m}] \sin m \psirp
\right.
\\&&
\phantom{\sum_{m\ne n}^{\infty}}
+\left.\left.
[\lambda^{s-}_{n+m} + \lambda^{s-}_{n-m}] \cos m \psirp
\right)
\right\}.
\end{eqnarray}
Here we have introduced the acceptance coefficient $a^{\pm}_{2n}$:
\begin{eqnarray}
\label{a_2n}
a_{2n}^{\pm} = \an{\pm}{n} = 1 \pm \ol{\cos 2n\phi},
\end{eqnarray}
and the following smallness parameters:
\begin{eqnarray}
\label{lambda_nm_u}
\lambda^{c\pm}_{m\mp n}  = \lc{m}{n}{\pm},~
\lambda^{s\pm}_{m\mp n}  = \ls{m}{n}{\pm}.
\end{eqnarray}
These parameters define the relative contribution of different terms
in Eq.~(\ref{x_n_PSI_acc_final}) and (\ref{y_n_PSI_acc_final}).
For a particular case of $m=n$,
values of $\lambda^{c,s\pm}_{2n}$
are defined only by detector acceptance,
while in general they also depend on the ratio of anisotropic flow $v_m$ and $v_n$.
For a perfect detector
$\overline{c}_m = \overline{s}_m = 0$
(and consequently all parameters $\lambda^{c,s\pm}_{m\mp n}=0$),
and Eqs.~(\ref{x_n_PSI_acc_final}) and (\ref{y_n_PSI_acc_final})
are reduced to Eq.~(\ref{x_n_PSI}).

Eqs.~(\ref{x_n_PSI_acc_final}) and (\ref{y_n_PSI_acc_final})
show that acceptance effects result in
coupling of equations for flow of different harmonics, and in general
a simultaneous analysis of all harmonics is required.
However we proceed below neglecting $m\ne n$ terms.
The relative contribution of $m\ne n$ terms
is of the order of $\lambda^{c,s\pm}_{n\mp m}$,
and this case can be understood as either when the $n$-th harmonic
flow is dominant: $v_n \gg v_{m\ne n}$
(such an assumption, for example, is made in \cite{Bhalerao:2003xf}
when discussing acceptance effects), or
the acceptance effects for higher harmonics are small:
$\overline{c},\overline{s}_{n\mp m}/a^{\pm}_{2n} \ll 1$, or both.

In the following, we distinguish three types of acceptance effects:
\begin{enumerate}
\item
{\it Shift of the $u_n$-vector}
due to non-zero values of $\overline{c}_n$ and $\overline{s}_n$
in Eqs.~(\ref{x_n_PSI_acc_final}) and (\ref{y_n_PSI_acc_final}).
This effect can be corrected for by subtracting
from the $u_n$-vector components
their corresponding average values:
\begin{eqnarray}
\label{u_recentering}
x'_n =  x_n - \overline{c}_n,~y'_n  =  y_n - \overline{s}_n.
\end{eqnarray}
This procedure is called {\it re-centering}.
\item
{\it Twist of the $u_n$-vector} 
that results in appearance of $\sin n
\psirp$ ($\cos n \psirp$) terms in $x_n$ ($y_n$) components of $u_n$-vector.
Determined by non-zero values of $\lambda^{s\pm}_{2n}$
in Eqs.~(\ref{x_n_PSI_acc_final}) and (\ref{y_n_PSI_acc_final}),
this effect can be corrected for by
the {\it diagonalization} procedure
(after re-centering has been applied):
\begin{eqnarray}
\label{u_rotation}
x''_n =  \frac{x'-\lambda^{s-}_{2n}y'}{1-\lambda^{s-}_{2n}\lambda^{s+}_{2n}},~
y''_n =  \frac{y'-\lambda^{s+}_{2n}x'}{1-\lambda^{s-}_{2n}\lambda^{s+}_{2n}}.
\end{eqnarray}
Twist effect is zero if $\overline{\sin 2 n \phi} =0$, e.g., in case $n=1$
for detectors symmetric in x or y (such as of rectangular shape).
\item
{\it Rescaling of the $u_n$-vector}, which is
defined by the coefficients $a^{\pm}_{2n}$
in Eqs.~(\ref{x_n_PSI_acc_final}) and (\ref{y_n_PSI_acc_final}).
This effect is the most important one next to the shift of
the $u_n$-vector, and it can be corrected for
by {\it rescaling} the
$u_n$-vector components with acceptance coefficients $a^{\pm}_{2n}$
(after the re-centering and twist corrections have been applied):
\begin{eqnarray}
\label{u_rescaling}
x'''_n =  \frac{x''_n}{a^{+}_{2n}},~y'''_n =  \frac{y''_n}{a^{-}_{2n}}.
\end{eqnarray}
\end{enumerate}
The acceptance corrected $u'''_n$-vector has the
same average $\mean{u'''_n}_{\psirp}$
as in the case of a detector with perfect acceptance
(compare with Eq.~(\ref{x_n_PSI})):
\begin{eqnarray}
\label{x_n_PSI_accCorrected}
\mean{u'''_n}_{\psirp} &=& v_n (\cos n \psirp + i \sin n \psirp).
\end{eqnarray}
Similar corrections can be applied
for the $Q_n$-vector components,
which we write in the following form
(contributions from $m\ne n$ terms have been neglected):
\begin{eqnarray}
\label{Q_acc}
\nonumber
\mean{X_n}_{\psirp} &=& \overline{X}_n+
A^{+}_{2n}
\left( \cos n \psirp + \Lambda^{s+}_{2n}\sin n \psirp \right),
\\
\nonumber
\mean{Y_n}_{\psirp} &=&\overline{Y}_n+
 A^{-}_{2n}
\left( \sin n \psirp + \Lambda^{s-}_{2n}\cos n \psirp \right).
\end{eqnarray}
The symmetry requires
$A^{+}_{2n}\Lambda^{s+}_{2n} = A^{-}_{2n}\Lambda^{s-}_{2n}$.
Applying corrections (\ref{u_recentering}-\ref{u_rescaling})
for the $Q_n$-vector
($a^\pm_{2n}$, $\lambda^\pm_{2n}$ have to be replaced with
$A^\pm_{2n}$, $\Lambda^\pm_{2n}$), one gets:
\begin{eqnarray}
\label{Q_acc_Corrected}
\mean{Q'''_n}_{\psirp} &=& \cos n \psirp + i\sin n \psirp.
\end{eqnarray}
From Eqs.~(\ref{x_n_PSI_accCorrected}) and (\ref{Q_acc_Corrected})
it follows that all equations given in section \ref{MethodReview}
can be applied to $u'''_n$ and $Q'''_n$ vectors,
and the same observables (\ref{XYseparetely}),
(\ref{mixedObsXY}), and (\ref{MixedHarmonicXYseparetely})
can be used for acceptance corrected $u'''_n$ and $Q'''_n$ vector components.

To clarify better how the above described corrections work
we consider below the correlations between uncorrected
$u_n$ and $Q_n$ vector components,
and discuss what kind of effects are removed by
each of the corrections (\ref{u_recentering}-\ref{u_rescaling}).


\subsection{Two particle correlations}

Acceptance effects in conjunction with anisotropic flow
may lead to various spurious correlations,
such as correlations in multiplicity 
and/or transverse momentum.
In particular, multiplicity correlations in two kinematic regions $a$ and $b$
are given by the following equation:
\begin{eqnarray}
\label{multiplicityCorr}
[\rho_a \rho_b] &\equiv& \int \frac{d\psirp}{2\pi}
~d\phi_a d\phi_b A(\phi_a) A(\phi_b)\rho_a \rho_b
\nonumber \\
&=& 1 + 2\sum_{m=1}^{\infty}
\left|\mean{v_m  \overline{u}_m}_{a} \mean{v_m \overline{u}^*_m}_{b} \right|,
\end{eqnarray}
where $\overline{u}_m= \overline{c}_m + i \overline{s}_m$,
and $\mean{...}_{a,b}$ denotes the average over 
kinematic regions $a$ and $b$.
According to Eqs.~(\ref{f_PSI_acc},~\ref{multiplicityCorr}),
the (event) average of the product of two functions $f$ and $g$
defined in regions $a$ and $b$ can be written as:
\begin{eqnarray}
\label{meanxXdefinition}
\mean{f_a g_b}&=& \frac{1}{[\rho_a \rho_b]}
\int \frac{d\psirp}{2\pi} \mean{f_a}_{\psirp} \mean{g_b}_{\psirp}.
\end{eqnarray}
Deviation of the denominator from unity
(which is the value for no multiplicity correlation)
is defined by non-zero terms in the sum 
in Eq.~(\ref{multiplicityCorr}) over $m$-harmonics.
Taking into account that the measured anisotropic flow at RHIC is $v_m \leq 10\%$,
in the most pessimistic estimate, using $ \mean{\overline{u}_m}_{a,b} \sim 1$,
we obtain $| [\rho_a \rho_b]-1| \leq 2\%$.
In practice, $\mean{\overline{u}_m}_{a,b} \leq 0.2 \div 0.3$,
which reduces the acceptance effects on multiplicity correlations
to the level of a tenth of a percent.
In principle, such effects
can be consistently taken into account, but
for the sake of simplicity,
below we proceed neglecting these multiplicity correlations.

In analogy to (\ref{meanDefin}) we consider
the correlations between uncorrected $u_n$
and $Q_n$ vector components:
\begin{eqnarray}
\label{xXCorrelationsAcc_vnOnly}
\nonumber
\mean{x_n X_n}&=&
\overline{x}_{n}\overline{X}_{n}+
\frac{v_n}{2}
a^+_{2n}A^+_{2n}
\left(1+\lambda^{s+}_{2n}\Lambda^{s+}_{2n}\right),
\\
\mean{y_n Y_n} &=&
\overline{y}_{n}\overline{Y}_{n}+
\frac{v_n}{2}a^-_{2n}A^-_{2n}
\left(1+\lambda^{s-}_{2n}\Lambda^{s-}_{2n}\right).
\end{eqnarray}

The first terms, $\overline{c}_n \overline{X}_{n}$ and $\overline{s}_n \overline{Y}_{n}$,
can be removed by the re-centering procedure (\ref{u_recentering}).
Note, that $\overline{c}_n$ and $\overline{X}_n$ enter as a product,
what allows to re-center only the event vector components.
Similar, to remove the second order terms, $\lambda^{s\pm}_{2n}\Lambda^{s\pm}_{2n}$,
it is sufficient to apply the twist correction only for the $Q_n$-vector components.
Parameters $A^\pm_{2n}$, and $\Lambda^{s\pm}_{2n}$
can be obtained with the random subevent technique.
In that case, they are defined by a set of coupled equations:
\begin{eqnarray}
\label{resolutionAcc_vnOnlyX}
8\mean{X'^a_n X'^b_n}
&=& {A^+_{2n}}^2 \left(1+{\Lambda^{s+}_{2n}}^2\right),
\\
\nonumber
8\mean{Y'^a_n Y'^b_n}
&=& {A^-_{2n}}^2 \left(1+{\Lambda^{s-}_{2n}}^2\right),
\\
\nonumber
8\mean{X'^a_n Y'^b_n}
&=&
A^+_{2n}A^-_{2n}  \left(\Lambda^{s+}_{2n}+\Lambda^{s-}_{2n}\right).
\end{eqnarray}
After the re-centering and twist corrections have been applied
Eqs.~(\ref{xXCorrelationsAcc_vnOnly}) leads to the following observable
for the anisotropic flow $v_n$:
\begin{eqnarray}
\label{v_nXAcc_vnOnly}
v_n = \frac{1}{a^+_{2n}}
\frac{\mean{x_n X''_n}}
{\sqrt{2 \mean{{X''^a_n} {X''^b_n}}}}
=
\frac{1}{a^-_{2n}}
\frac{\mean{y_n Y''_n}}
{\sqrt{2 \mean{{Y''^a_n} {Y''^b_n}}}}.
\end{eqnarray}
Rescaling of the $u_n$-vector reduces this to Eq.~(\ref{XYseparetely}),
which should be written for rescaled $u_n$ and shifted and twisted $Q''_n$ vectors.
Note, that correction factors $\sqrt{\mean{{X''^a_n} {X''^b_n}}}$
and $\sqrt{\mean{{Y''^a_n} {Y''^b_n}}}$ correspond
to rescaling of the $Q''_n$-vector.

Eq.~(\ref{v_nXAcc_vnOnly}) shows that
in case of two particle correlations
re-centering and twist corrections of the $u_n$-vector are not required.
This equation can be also used for the $Q'_n$-vector,
if the second order corrections defined by the terms
$\lambda^{s\pm}_{2n}\Lambda^{s\pm}_{2n}$
are small, and can be neglected (twist correction is not required).


\subsection{Three particle correlations}
In the case of three particle correlations we consider:
\begin{eqnarray}
\label{xxXCorrelationsAcc}
\nonumber
\mean{x^a_n x^b_n X_{2n}} &=&
\mean{x^a_n x^b_n}\overline{X}_{2n}+
\overline{x^a_n}\mean{x^b_n X_{2n}}+
\overline{x^b_n}\mean{x^a_n X_{2n}}
\\
&&+\frac{v_n^2}{4} {a^+_{2n}}^2 A^+_{4n}
\left(1-{\lambda^{s+}_{2n}}^2 +2\lambda^{s+}_{2n}\Lambda^{s+}_{4n}\right).
\end{eqnarray}
Similar expressions can be written for other terms in Eq.~(\ref{mixedObsXY}).
In contrast to the case of two particle correlations,
re-centering procedure (\ref{u_recentering}) is required for all
three vectors $u_n^a$, $u_n^b$, and $Q_{2n}$.
Acceptance coefficients $A^{\pm}_{4n}$ and $\Lambda^{s\pm}_{4n}$
are given by a set of coupled equations (\ref{resolutionAcc_vnOnlyX})
written for the $Q_{2n}$-vector.
Twist corrections applied to  $Q_{2n}$ and $u^{a,b}_n$ vectors removes
the terms ${\lambda^{s\pm}_{2n}}^2$ and $2\lambda^{s\pm}_{2n}\Lambda^{s\pm}_{4n}$,
and Eq.~(\ref{xxXCorrelationsAcc}) leads to observable (\ref{MixedHarmonicXYseparetely})
written for $u'''_n$ and $Q'''_n$ vector components.


\section{\label{Conclusion} Method summary and conclusion} 

In this paper we discuss new observables for anisotropic flow measurement
based on correlations of $x$ and $y$  components of the flow vectors.
Providing independent measures of anisotropic flow they can be used
to check self-consistency of the analysis.
Moreover, these observables allow direct accounting
for acceptance effects,
which we discuss in detail for two particular cases of anisotropic flow measurement
with two and three (mixed harmonic) particle correlations.
Importantly, acceptance corrections and parameters, which define applicability 
of these observables,
can be determined directly from experimental data.

Non-uniformity of the detector acceptance is
quantified with coefficients $\cn{n} =\ol{\cos n\phi}$ and 
$\sn{n}=\ol{\sin n\phi}$,
which further define coefficients $a_{2n}^{\pm}$ and $\lambda^{c,s\mp}_{m\pm n}$
given by Eqs.~(\ref{a_2n}) and (\ref{lambda_nm_u}).
Though accounting for the  acceptance effects in general might be
difficult as it requires a solution of a set of coupled equations
with different harmonics involved,
in the case when the
contribution of $m \ne n$ harmonics can be neglected,
the problem significantly simplifies.
It becomes possible to correct
the single particle $u_n$ and event flow $Q_n$ vectors
such that the conventional observables (derived for the 
perfect detector) can be used.
Note, that both, the acceptance
coefficients  $\cn{n}$ and $\sn{n}$, and the correlators between 
$u_n$ and $Q_n$ vectors can be obtained during
a single pass over the data.
This can significantly reduce the amount of time needed for the calculation.
At the same time, due to variation of detector acceptance in time,
with collision centrality, vertex position, etc.,
it may be important to apply the acceptance corrections
separately run-by-run, for different vertex position, etc.
In that case it might be more convenient to split the procedure
into a few steps with two passes over the data.
During the first pass 
the acceptance coefficients $\cn{n}$
and $\sn{n}$ are extracted
as a function of different centrality, time, etc.
and all coefficients needed for acceptance corrections
presented in
Eqs.~(\ref{u_recentering}-\ref{u_rescaling}), both, for $u_n$ and
$Q_n$ vectors, are calculated.
During the second pass over the data the correlators of the standard 
procedure given by Eq.~(\ref{ScalarObsXY}-\ref{ScalarResXY})
(Eq.~(\ref{mixedObsXY}) in the case of three particle
correlations) are calculated. Finally, the flow values are extracted
as given in Eqs.~(\ref{XYseparetely}), (\ref{MixedHarmonicXYseparetely}).

\section*{Acknowledgments}
We thank the members of the STAR flow physics discussion group for fruitful
discussions, and especially A.M.~Poskanzer and C. Pruneau
for reading the manuscript and fruitful suggestions, and J.-Y. Ollitrault
for critical comments.
Financial support provided in part by US Department
of Energy Grant No. DE-FG02-92ER40713.

\bibliography{flowAsymmetricAcceptance}

\begin{thebibliography}{31}
\expandafter\ifx\csname natexlab\endcsname\relax\def\natexlab#1{#1}\fi
\expandafter\ifx\csname bibnamefont\endcsname\relax
  \def\bibnamefont#1{#1}\fi
\expandafter\ifx\csname bibfnamefont\endcsname\relax
  \def\bibfnamefont#1{#1}\fi
\expandafter\ifx\csname citenamefont\endcsname\relax
  \def\citenamefont#1{#1}\fi
\expandafter\ifx\csname url\endcsname\relax
  \def\url#1{\texttt{#1}}\fi
\expandafter\ifx\csname urlprefix\endcsname\relax\def\urlprefix{URL }\fi
\providecommand{\bibinfo}[2]{#2}
\providecommand{\eprint}[2][]{\url{#2}}

\bibitem[{\citenamefont{Back et~al.}(2005{\natexlab{a}})}]{Back:2004je}
\bibinfo{author}{\bibfnamefont{B.~B.} \bibnamefont{Back}} \bibnamefont{et~al.},
  \bibinfo{journal}{Nucl. Phys.} \textbf{\bibinfo{volume}{A757}},
  \bibinfo{pages}{28} (\bibinfo{year}{2005}{\natexlab{a}}),
  \eprint{nucl-ex/0410022}.

\bibitem[{\citenamefont{Adams et~al.}(2005{\natexlab{a}})}]{Adams:2005dq}
\bibinfo{author}{\bibfnamefont{J.}~\bibnamefont{Adams}} \bibnamefont{et~al.}
  (\bibinfo{collaboration}{STAR}), \bibinfo{journal}{Nucl. Phys.}
  \textbf{\bibinfo{volume}{A757}}, \bibinfo{pages}{102}
  (\bibinfo{year}{2005}{\natexlab{a}}), \eprint{nucl-ex/0501009}.

\bibitem[{\citenamefont{Adcox et~al.}(2005)}]{Adcox:2004mh}
\bibinfo{author}{\bibfnamefont{K.}~\bibnamefont{Adcox}} \bibnamefont{et~al.}
  (\bibinfo{collaboration}{PHENIX}), \bibinfo{journal}{Nucl. Phys.}
  \textbf{\bibinfo{volume}{A757}}, \bibinfo{pages}{184} (\bibinfo{year}{2005}),
  \eprint{nucl-ex/0410003}.

\bibitem[{\citenamefont{Voloshin}(2003)}]{Voloshin:2002wa}
\bibinfo{author}{\bibfnamefont{S.~A.} \bibnamefont{Voloshin}},
  \bibinfo{journal}{Nucl. Phys.} \textbf{\bibinfo{volume}{A715}},
  \bibinfo{pages}{379} (\bibinfo{year}{2003}), \eprint{nucl-ex/0210014}.

\bibitem[{\citenamefont{Voloshin and Zhang}(1996)}]{Voloshin:1994mz}
\bibinfo{author}{\bibfnamefont{S.}~\bibnamefont{Voloshin}} \bibnamefont{and}
  \bibinfo{author}{\bibfnamefont{Y.}~\bibnamefont{Zhang}}, \bibinfo{journal}{Z.
  Phys.} \textbf{\bibinfo{volume}{C70}}, \bibinfo{pages}{665}
  (\bibinfo{year}{1996}), \eprint{hep-ph/9407282}.

\bibitem[{\citenamefont{Poskanzer and
  Voloshin}(1998{\natexlab{a}})}]{Poskanzer:1998yz}
\bibinfo{author}{\bibfnamefont{A.~M.} \bibnamefont{Poskanzer}}
  \bibnamefont{and} \bibinfo{author}{\bibfnamefont{S.~A.}
  \bibnamefont{Voloshin}}, \bibinfo{journal}{Phys. Rev.}
  \textbf{\bibinfo{volume}{C58}}, \bibinfo{pages}{1671}
  (\bibinfo{year}{1998}{\natexlab{a}}), \eprint{nucl-ex/9805001}.

\bibitem[{\citenamefont{Barrette et~al.}(1994)}]{Barrette:1994xr}
\bibinfo{author}{\bibfnamefont{J.}~\bibnamefont{Barrette}} \bibnamefont{et~al.}
  (\bibinfo{collaboration}{E877}), \bibinfo{journal}{Phys. Rev. Lett.}
  \textbf{\bibinfo{volume}{73}}, \bibinfo{pages}{2532} (\bibinfo{year}{1994}),
  \eprint{hep-ex/9405003}.

\bibitem[{\citenamefont{Barrette et~al.}(1997)}]{Barrette:1996rs}
\bibinfo{author}{\bibfnamefont{J.}~\bibnamefont{Barrette}} \bibnamefont{et~al.}
  (\bibinfo{collaboration}{E877}), \bibinfo{journal}{Phys. Rev.}
  \textbf{\bibinfo{volume}{C55}}, \bibinfo{pages}{1420} (\bibinfo{year}{1997}),
  \eprint{nucl-ex/9610006}.

\bibitem[{\citenamefont{Appelshauser et~al.}(1998)}]{Appelshauser:1997dg}
\bibinfo{author}{\bibfnamefont{H.}~\bibnamefont{Appelshauser}}
  \bibnamefont{et~al.} (\bibinfo{collaboration}{NA49}), \bibinfo{journal}{Phys.
  Rev. Lett.} \textbf{\bibinfo{volume}{80}}, \bibinfo{pages}{4136}
  (\bibinfo{year}{1998}), \eprint{nucl-ex/9711001}.

\bibitem[{\citenamefont{Alt et~al.}(2003)}]{Alt:2003ab}
\bibinfo{author}{\bibfnamefont{C.}~\bibnamefont{Alt}} \bibnamefont{et~al.}
  (\bibinfo{collaboration}{NA49}), \bibinfo{journal}{Phys. Rev.}
  \textbf{\bibinfo{volume}{C68}}, \bibinfo{pages}{034903}
  (\bibinfo{year}{2003}), \eprint{nucl-ex/0303001}.

\bibitem[{\citenamefont{Ackermann et~al.}(2001)}]{Ackermann:2000tr}
\bibinfo{author}{\bibfnamefont{K.~H.} \bibnamefont{Ackermann}}
  \bibnamefont{et~al.} (\bibinfo{collaboration}{STAR}), \bibinfo{journal}{Phys.
  Rev. Lett.} \textbf{\bibinfo{volume}{86}}, \bibinfo{pages}{402}
  (\bibinfo{year}{2001}), \eprint{nucl-ex/0009011}.

\bibitem[{\citenamefont{Adler et~al.}(2001)}]{Adler:2001nb}
\bibinfo{author}{\bibfnamefont{C.}~\bibnamefont{Adler}} \bibnamefont{et~al.}
  (\bibinfo{collaboration}{STAR}), \bibinfo{journal}{Phys. Rev. Lett.}
  \textbf{\bibinfo{volume}{87}}, \bibinfo{pages}{182301}
  (\bibinfo{year}{2001}), \eprint{nucl-ex/0107003}.

\bibitem[{\citenamefont{Adler et~al.}(2003)}]{Adler:2003kt}
\bibinfo{author}{\bibfnamefont{S.~S.} \bibnamefont{Adler}} \bibnamefont{et~al.}
  (\bibinfo{collaboration}{PHENIX}), \bibinfo{journal}{Phys. Rev. Lett.}
  \textbf{\bibinfo{volume}{91}}, \bibinfo{pages}{182301}
  (\bibinfo{year}{2003}), \eprint{nucl-ex/0305013}.

\bibitem[{\citenamefont{Adcox et~al.}(2002)}]{Adcox:2002ms}
\bibinfo{author}{\bibfnamefont{K.}~\bibnamefont{Adcox}} \bibnamefont{et~al.}
  (\bibinfo{collaboration}{PHENIX}), \bibinfo{journal}{Phys. Rev. Lett.}
  \textbf{\bibinfo{volume}{89}}, \bibinfo{pages}{212301}
  (\bibinfo{year}{2002}), \eprint{nucl-ex/0204005}.

\bibitem[{\citenamefont{Back et~al.}(2002)}]{Back:2002gz}
\bibinfo{author}{\bibfnamefont{B.~B.} \bibnamefont{Back}} \bibnamefont{et~al.}
  (\bibinfo{collaboration}{PHOBOS}), \bibinfo{journal}{Phys. Rev. Lett.}
  \textbf{\bibinfo{volume}{89}}, \bibinfo{pages}{222301}
  (\bibinfo{year}{2002}), \eprint{nucl-ex/0205021}.

\bibitem[{\citenamefont{Back et~al.}(2005{\natexlab{b}})}]{Back:2004mh}
\bibinfo{author}{\bibfnamefont{B.~B.} \bibnamefont{Back}} \bibnamefont{et~al.}
  (\bibinfo{collaboration}{PHOBOS}), \bibinfo{journal}{Phys. Rev.}
  \textbf{\bibinfo{volume}{C72}}, \bibinfo{pages}{051901}
  (\bibinfo{year}{2005}{\natexlab{b}}), \eprint{nucl-ex/0407012}.

\bibitem[{\citenamefont{Back et~al.}(2005{\natexlab{c}})}]{Back:2004zg}
\bibinfo{author}{\bibfnamefont{B.~B.} \bibnamefont{Back}} \bibnamefont{et~al.}
  (\bibinfo{collaboration}{PHOBOS}), \bibinfo{journal}{Phys. Rev. Lett.}
  \textbf{\bibinfo{volume}{94}}, \bibinfo{pages}{122303}
  (\bibinfo{year}{2005}{\natexlab{c}}), \eprint{nucl-ex/0406021}.

\bibitem[{\citenamefont{Ollitrault}(1992)}]{Ollitrault:1992bk}
\bibinfo{author}{\bibfnamefont{J.-Y.} \bibnamefont{Ollitrault}},
  \bibinfo{journal}{Phys. Rev.} \textbf{\bibinfo{volume}{D46}},
  \bibinfo{pages}{229} (\bibinfo{year}{1992}).

\bibitem[{\citenamefont{Teaney et~al.}(2002)\citenamefont{Teaney, Lauret, and
  Shuryak}}]{Teaney:2001gc}
\bibinfo{author}{\bibfnamefont{D.}~\bibnamefont{Teaney}},
  \bibinfo{author}{\bibfnamefont{J.}~\bibnamefont{Lauret}}, \bibnamefont{and}
  \bibinfo{author}{\bibfnamefont{E.~V.} \bibnamefont{Shuryak}},
  \bibinfo{journal}{Nucl. Phys.} \textbf{\bibinfo{volume}{A698}},
  \bibinfo{pages}{479} (\bibinfo{year}{2002}), \eprint{nucl-th/0104041}.

\bibitem[{\citenamefont{Teaney et~al.}(2001)\citenamefont{Teaney, Lauret, and
  Shuryak}}]{Teaney:2001av}
\bibinfo{author}{\bibfnamefont{D.}~\bibnamefont{Teaney}},
  \bibinfo{author}{\bibfnamefont{J.}~\bibnamefont{Lauret}}, \bibnamefont{and}
  \bibinfo{author}{\bibfnamefont{E.~V.} \bibnamefont{Shuryak}}
  (\bibinfo{year}{2001}), \eprint{nucl-th/0110037}.

\bibitem[{\citenamefont{Poskanzer and
  Voloshin}(1998{\natexlab{b}})}]{Poskanzer:1998lbnl}
\bibinfo{author}{\bibfnamefont{A.}~\bibnamefont{Poskanzer}} \bibnamefont{and}
  \bibinfo{author}{\bibfnamefont{S.}~\bibnamefont{Voloshin}},
  \bibinfo{journal}{LBNL Annual Report}  (\bibinfo{year}{1998}{\natexlab{b}}),
  \eprint{http://ie.lbl.gov/nsd1999/rnc/RNC.htm}.

\bibitem[{\citenamefont{Adler et~al.}(2002)}]{Adler:2002pu}
\bibinfo{author}{\bibfnamefont{C.}~\bibnamefont{Adler}} \bibnamefont{et~al.}
  (\bibinfo{collaboration}{STAR}), \bibinfo{journal}{Phys. Rev.}
  \textbf{\bibinfo{volume}{C66}}, \bibinfo{pages}{034904}
  (\bibinfo{year}{2002}), \eprint{nucl-ex/0206001}.

\bibitem[{\citenamefont{Borghini et~al.}(2002)\citenamefont{Borghini, Dinh, and
  Ollitrault}}]{Borghini:2002vp}
\bibinfo{author}{\bibfnamefont{N.}~\bibnamefont{Borghini}},
  \bibinfo{author}{\bibfnamefont{P.~M.} \bibnamefont{Dinh}}, \bibnamefont{and}
  \bibinfo{author}{\bibfnamefont{J.~Y.} \bibnamefont{Ollitrault}},
  \bibinfo{journal}{Phys. Rev.} \textbf{\bibinfo{volume}{C66}},
  \bibinfo{pages}{014905} (\bibinfo{year}{2002}), \eprint{nucl-th/0204017}.

\bibitem[{\citenamefont{Adams et~al.}(2004)}]{Adams:2003zg}
\bibinfo{author}{\bibfnamefont{J.}~\bibnamefont{Adams}} \bibnamefont{et~al.}
  (\bibinfo{collaboration}{STAR}), \bibinfo{journal}{Phys. Rev. Lett.}
  \textbf{\bibinfo{volume}{92}}, \bibinfo{pages}{062301}
  (\bibinfo{year}{2004}), \eprint{nucl-ex/0310029}.

\bibitem[{\citenamefont{Adams et~al.}(2005{\natexlab{b}})}]{Adams:2004bi}
\bibinfo{author}{\bibfnamefont{J.}~\bibnamefont{Adams}} \bibnamefont{et~al.}
  (\bibinfo{collaboration}{STAR}), \bibinfo{journal}{Phys. Rev.}
  \textbf{\bibinfo{volume}{C72}}, \bibinfo{pages}{014904}
  (\bibinfo{year}{2005}{\natexlab{b}}), \eprint{nucl-ex/0409033}.

\bibitem[{\citenamefont{Aizawa et~al.}(2003)}]{Aizawa:2003zq}
\bibinfo{author}{\bibfnamefont{M.}~\bibnamefont{Aizawa}} \bibnamefont{et~al.}
  (\bibinfo{collaboration}{PHENIX}), \bibinfo{journal}{Nucl. Instrum. Meth.}
  \textbf{\bibinfo{volume}{A499}}, \bibinfo{pages}{508} (\bibinfo{year}{2003}).

\bibitem[{\citenamefont{Adcox et~al.}(2003)}]{Adcox:2003zp}
\bibinfo{author}{\bibfnamefont{K.}~\bibnamefont{Adcox}} \bibnamefont{et~al.}
  (\bibinfo{collaboration}{PHENIX}), \bibinfo{journal}{Nucl. Instrum. Meth.}
  \textbf{\bibinfo{volume}{A499}}, \bibinfo{pages}{489} (\bibinfo{year}{2003}).

\bibitem[{\citenamefont{Afanasev et~al.}(1999)}]{Afanasev:1999iu}
\bibinfo{author}{\bibfnamefont{S.}~\bibnamefont{Afanasev}} \bibnamefont{et~al.}
  (\bibinfo{collaboration}{NA49}), \bibinfo{journal}{Nucl. Instrum. Meth.}
  \textbf{\bibinfo{volume}{A430}}, \bibinfo{pages}{210} (\bibinfo{year}{1999}).

\bibitem[{\citenamefont{\mbox{}STAR ZDC-SMD~proposal}(2003)}]{ZDCSMDproposal}
\bibinfo{author}{\bibnamefont{\mbox{}STAR ZDC-SMD~proposal}}
  (\bibinfo{collaboration}{STAR}), \bibinfo{journal}{STAR Note}
  \textbf{\bibinfo{volume}{SN0448}} (\bibinfo{year}{2003}).

\bibitem[{\citenamefont{Borghini et~al.}(2001)\citenamefont{Borghini, Dinh, and
  Ollitrault}}]{Borghini:2001vi}
\bibinfo{author}{\bibfnamefont{N.}~\bibnamefont{Borghini}},
  \bibinfo{author}{\bibfnamefont{P.~M.} \bibnamefont{Dinh}}, \bibnamefont{and}
  \bibinfo{author}{\bibfnamefont{J.-Y.} \bibnamefont{Ollitrault}},
  \bibinfo{journal}{Phys. Rev.} \textbf{\bibinfo{volume}{C64}},
  \bibinfo{pages}{054901} (\bibinfo{year}{2001}), \eprint{nucl-th/0105040}.

\bibitem[{\citenamefont{Bhalerao et~al.}(2003)\citenamefont{Bhalerao, Borghini,
  and Ollitrault}}]{Bhalerao:2003xf}
\bibinfo{author}{\bibfnamefont{R.~S.} \bibnamefont{Bhalerao}},
  \bibinfo{author}{\bibfnamefont{N.}~\bibnamefont{Borghini}}, \bibnamefont{and}
  \bibinfo{author}{\bibfnamefont{J.~Y.} \bibnamefont{Ollitrault}},
  \bibinfo{journal}{Nucl. Phys.} \textbf{\bibinfo{volume}{A727}},
  \bibinfo{pages}{373} (\bibinfo{year}{2003}), \eprint{nucl-th/0310016}.

\end{thebibliography}

\end{document}